\documentstyle[preprint,prl,aps]{revtex}
\begin{document}
\preprint{Preprint Version No. 3.0}
\draft
\title{Detailed Structure of a CDW in a Quenched Random Field}
\author{J.D.~Brock,
A.C.~Finnefrock,
K.L.~Ringland,
and E.~Sweetland}
\address{School of Applied and Engineering Physics \\
Cornell University, Ithaca, New York  14853}

\date{\today}
\maketitle
\begin{abstract}

Using high resolution x-ray scattering, we have measured
the structure of the ${\bf Q}_1$ CDW in Ta-doped NbSe$_3$.
Detailed line shape analysis of the
data demonstrates that two length scales are required to describe
the phase-phase correlation function.
Phase fluctuations with
wavelengths less than a new length scale $a$ are suppressed
and this $a$ is identified with the amplitude
coherence length.
We find that
$\xi_{{\bf a}^*} = 34.4 \pm 10.3$\AA\@.
Implications for the physical mechanisms responsible for
pinning are discussed.
\end{abstract}


During the past twenty years,
the influence of random disorder on phase transitions
has been studied extensively.
Disorder can be formally described as
randomness in either the interaction strength\cite{harris}
or the field conjugate to the order parameter\cite{imry}.
It is widely believed that a small amount of randomness in the
interactions is irrelevant.
Random fields, on the other hand, have dramatic
consequences.
In their seminal paper, Imry and Ma\cite{imry} suggested that
a random field should cause the lower marginal dimensionality
$d_{\ell}$ to rise from 2 to 4 for systems with continuous
symmetry.
Here, $d_{\ell}$ is the dimensionality below which the system
cannot sustain long-range order (LRO) at finite temperatures\cite{landau}.
This loss of LRO has been observed in a wide variety of
systems.
In particular, the structures of the charge-density waves (CDW's) found in
the quasi-one-dimensional metal NbSe$_3$ do not exhibit
LRO\cite{sweetland,daved}.

For mathematical simplicity,
the quenched random field is frequently assumed to be
a time-independent random Gaussian variable satisfying
$\left < h ( {\bf r }) \right> = 0$
and
$\left < h ( {\bf r}) h ( {\bf r }^{\prime} ) \right > =
n_i h_0^2 \ \delta^d ({\bf r} - {\bf r}^{\prime} ) $
where
$n_i$ is the number density of impurity atoms and
$h_0$ gives the strength of the pinning interaction.
In the general case, scattering from a random structure
is characterized by an exponentially decaying correlation
function\cite{debye}.
For the specific case of CDW systems,
one expects an exponential decay of the correlations between
static fluctuations in the phase of the CDW order parameter\cite{sham,efetov}.

In this paper, we report the results of a detailed high resolution x-ray
scattering study of the static phase correlations
of the ${\bf Q}_1$ CDW in Ta-doped NbSe$_3$.
The experiments clearly show that phase fluctuations with
high spatial frequency components are suppressed;
therefore,
the destruction of LRO by
the random field is suppressed on length scales less than
$\approx 75$~\AA.
These results are explained using the standard Ginzburg-Landau
phase Hamiltonian with a Gaussian random pinning field\cite{sham,efetov}.
This new length scale is then related directly to the amplitude coherence
length.
We conclude with a brief discussion of the implications for
physical models of the pinning mechanism.

The obvious way to study the interaction between a CDW and
a quenched random distribution of impurity atoms is to dope
crystals with impurities and study how the ordering varies
with impurity concentration and type.
Over the last decade a large number of such studies have been
reported\cite{REFS,thorne}.
Most of these studies have focussed on NbSe$_3$,
which undergoes two independent
Peierls transitions to CDW states at $T_{P_1} \approx 145$~K and
at $T_{P_2} \approx 59$~K.
The most widely studied dopant has been Ta, which is isoelectronic
with Nb.

Experimentally\cite{sweetland,daved},
the structure of the ${\bf Q}_1$ CDW in Ta-doped
NbSe$_3$ is quite well
described by a Ginzburg-Landau field theory\cite{FLR}.
The effective Hamiltonian which describes the phase behavior
of the ${\bf Q}_1$ CDW at low temperatures can be written as\cite{sham}
\begin{equation}
{\cal H}_{\phi} = \int d^d {\bf x}
\left \{
\psi^2 \left (
{\mbox {\boldmath $\xi$}}
\cdot {\bf \nabla} \phi ( {\bf x} ) \right )^2
+ \psi  h ({\bf x}) \phi ({\bf x })
\right \} ,
\end{equation}
where
$\psi$ is the amplitude of CDW order parameter,
$\phi({\bf x})$ is the phase of the CDW order parameter,
{\boldmath $\xi$} is the amplitude coherence length, and
$h({\bf x})$
is the quenched random field.

Using ${\cal H}_{\phi}$
and the assumption that $h({\bf x})$ obeys
Gaussian statistics,
the phase-phase correlation function in three dimensions
is\cite{sham,efetov}
$
\left < e^{i[ \phi ({\bf x}_1) - \phi({\bf x}_2)]} \right >
\ \sim \
e^{-g({\bf x}_1 - {\bf x}_2 )}
$,
where
\begin{equation}
g ({\bf x})
=
\int_{k=1/L}^{k=1/\xi} \frac {d^3 k}
{(2 \pi )^3}
\ \frac { n_i h_0^2 } { 4 \xi^4 \psi^2 k^4 }
\ \left\{
1 - \cos ( {\bf k} \cdot {\bf x} )
\right\} ,
\label{eq:chi}
\end{equation}
and $L$ is the system size.
At large separations,
$\left < e^{i[ \phi ({\bf x}_1) - \phi({\bf x}_2)]}
\right > \sim e^{-|{\bf x}_1 - {\bf x}_2 |/\ell}$,
where
the correlation length is given by
$\ell = \frac { 32 \pi \xi^4 \psi^2}{n_i h_0^2} $.

In order to provide a framework for discussing the experimental results,
we first consider the consequences of random fields
on the measured profiles.
In an x-ray scattering experiment on a CDW system at low temperatures, the
static structure factor ${\cal S}({\bf q})$
is proportional to the spatial Fourier transform
of the phase-phase correlation function\cite{daved}.
Thus,
at low temperatures ${\cal S}({\bf q})$ exhibits
Lorentzian squared fluctuations centered about the
CDW satellite reflection positions.

The experimentally measured profile is the convolution of
${\cal S}({\bf q})$ with the resolution function of the
diffractometer, ${\cal R}$({\boldmath $\zeta$}).
If the sample has nontrivial mosaic structure, that must also be
convolved into the resolution function.
The form of ${\cal R}$({\boldmath $\zeta$}) can be calculated
from first principles\cite{james}.
To obtain a simple analytic form of the resolution function, we
assume
({\em i\/})
that the line shape of the rocking curves of the
monochromator and analyzer are Gaussians,
({\em ii\/})
that the half-widths of these curves are determined by the Darwin width
of Si(111),
({\em iii\/})
that the angular profile of the wiggler beam is Gaussian,
and
({\em iv\/})
that the angular divergence of the wiggler beam is that same
as that of a bend magnet at the critical wave length.
The assumed Gaussian line shapes allow the integrals to be done
analytically.
The result is then transformed from angle space to
reciprocal space,
and fit to an anisotropic Gaussian line shape.
The half widths are
$W_{\parallel}  = 2.08 \times 10^{-4}$\AA$^{-1}$ and
$W_{\perp} = 2.06 \times 10^{-5}$\AA$^{-1}$.
The long axis is rotated by $7.48^\circ$ from
the longitudinal direction.
The ellipse describing the half height of $R(${\boldmath$ \zeta$}$)$
is plotted in
the inset to Fig.~\ref{fig:pure_b}.
We will show below that
the intrinsic scattering is, in fact, much
narrower than the resolution of the diffractometer in both the
$q_{\parallel}$ (longitudinal) and the $q_z$ (out of scattering plane)
directions.
Therefore,
the directions orthogonal to a transverse scan
are integrated away and the dimensionality of the
Fourier transform
is effectively lowered from three to one\cite{daved}.
Thus, the observed line shape is given by the one-dimensional
convolution of the intrinsic scattering with
the resolution function.
Since
the one-dimensional Fourier transform of an exponentially
decaying correlation function is a simple Lorentzian,
the observed scattering will be the convolution of a Gaussian
resolution function
and a Lorentzian structure factor.

The experiments were carried out using a six-circle
diffractometer on the X-25 wiggler beam line at the National Synchrotron
Light Source.
The beam line optics include a toroidally focusing mirror,
a double crystal Si(111) monochromator,
and a triple-bounce channel-cut Si(111) analyzer crystal.
Non-dispersive rocking curves of the analyzer produced Darwin-limited
behavior.
Our samples were single-crystal whiskers of NbSe$_3$ lightly
doped with Ta
$\left ( \frac {R(300\:{\rm K})}{R(4\:{\rm K})} \equiv r_R \approx 45 \right
)$.
The macroscopic sample dimensions were approximately
2~$\mu$m $\times$
20~$\mu$m $\times$
5~mm.
The crystal structure of NbSe$_3$ is monoclinic and the
real space lattice constants\cite{hodeau} are
$a = 10.009$\AA,
$b = 3.4805$\AA,
$c = 15.629$\AA,
and $\beta = 109.47^{\circ}$.
We studied only the $T_{P_1}$ CDW whose wave vector is
${\bf Q}_1$~=~[0~$Q_1$~0], where $Q_1 \approx 0.243$
and varies slightly with temperature\cite{moudden}.
The samples were mounted using silver paint across a 3.5~mm hole
in an alumina substrate and studied in transmission.
Sample cooling was provided by a closed-cycle helium refrigerator.
All of the data presented in this paper were collected at 70~K.

Figure \ref{fig:pure_b}
shows a scan in the ${\bf b}^*$ direction
through the [0~$\overline{1+Q_1}$~0] CDW satellite
of sample~\#1.
During a typical injection cycle, the stored electron beam current
decayed from roughly 200 to 150 mA\@.
The data was collected for a constant number of incident x-rays and
is plotted as counts per sec assuming 100 mA in the storage
ring.
The half-width at half-maximum (HWHM) of the scan in the ${\bf b}^*$
direction is $1.2 \times 10^{-4}$\AA$^{-1}$, which is 10\% smaller
than the value obtained from the cut through the
{\em a priori} resolution function appropriate for a resolution-limited peak.
This cut is
indicated by the arrows in the inset to Fig.~\ref{fig:pure_b}.
Thus,
as claimed above, this scan is
{\em completely} resolution limited.
Figures \ref{fig:pure_a}
and
\ref{fig:doped_a}
show scans through the [0~$\overline{1+Q_1}$~0] CDW satellite in the
${\bf a}^*$ direction for two different samples.
The Gaussian resolution function in the ${\bf a}^*$ direction,
which now includes the sample mosaic as measured at the
[0~$\overline{2}$~0] Bragg peak,
is indicated by the dot-dashed line in the
inset of each figure and the values of the HWHM, $W_{\rm R}$,
are listed in table \ref{tb:summary}.
The dashed lines in Figs.~\ref{fig:pure_a} and \ref{fig:doped_a}
are the best fit to
the convolution of a simple Lorentzian and the Gaussian resolution
for the two samples.
The background level I$_{\rm BG}$ has been constrained to remain at or above
the measured value of the background.
A small asymmetry factor has been included.
This factor improves the quality of the fit but does not
affect the values of the other parameters.
A casual inspection of Figs.~\ref{fig:pure_a} and \ref{fig:doped_a}
reveals two important points.
First,
the high brightness wiggler source associated with X-25
allowed us to obtain data over four
decades of intensity, roughly two decades better than in
previous measurements\cite{daved}.
Second, this extra dynamic range clearly reveals that
the simple Lorentzian line shape overestimates the scattering
in the wings.

It is not surprising that the tails do not agree with
a simple Lorentzian line shape.
The exponential decay of the phase-phase correlation function
is valid only at large separations, breaking down at small
separations.
The magnitude of
this discrepancy can be made quantitative by considering
the Taylor series expansion of Eq.~(\ref{eq:chi})
about $x = 0$.
\begin{equation}
g(x) \approx
\frac {2}{3 \pi \ell} \left ( \frac {1}{\xi} - \frac {1}{L} \right ) x^2
+
{\cal O} \left ( x^4 \right )
\label{eq:taylor}
\end{equation}
In contrast to the asymptotic form,
the full functional form of $g(x)$ approaches the origin with zero slope.

Previously,
high resolution x-ray scattering studies of the structure of
the Pt(001) surface\cite{abernathy} and high stage Br-intercalated
graphite\cite{mochrie} have observed that it is
necessary to approximate $g ({\bf x})$ in a manner which
preserves the {\em functional} form at both large
and small distances\cite{comment}.
In our particular case, this can be done by letting
$g(x) = \sqrt{ a^2 + x^2}/\ell$.
As before,
the observed line shape is
given by a one-dimensional convolution in reciprocal space
of the static structure factor
with the resolution function.
The one-dimensional static structure factor describing
transverse scans through the CDW satellite produced by this model is\cite{GR}
\begin{eqnarray}
{\cal S}(q_{\perp})
& \sim &
\int_{- \infty}^{\infty} dx e^{i q_{\perp} x}
\exp \left [ - \frac { \sqrt { a^2 ~ + x^2 } }
{\ell} \right ] \\
& = &
\frac { 2 a / \ell } {\sqrt{ \ell^{-2} + q_{\perp}^2}}
\;
K_1 \left ( a\sqrt{ \ell^{-2} + q_{\perp}^2} \ \right )
\label{eq:K_1}
\end{eqnarray}
where $K_1(x)$ is the modified Bessel function of order~1.
For small arguments, $K_1(x) \approx x^{-1} $ and
the simple Lorentzian form is recovered.
For large arguments, $K_1(x) \approx
\frac {\pi} {\sqrt{ 2 \pi x}} \exp (-x)$
and the scattering decays exponentially.
Qualitatively,
introducing a cutoff at small $x$ reduces the scattering in the wings.

The solid lines in Figs.~\ref{fig:pure_a} and \ref{fig:doped_a}
are the best fit to the convolution of Eq.~(\ref{eq:K_1}) with
the Gaussian resolution functions shown in the insets.
The agreement between the fits and the data is striking.
Quantitatively, for sample~\#1,
the goodness-of-fit parameter $\chi^2$ changes from
$26.7$ to $5.5$ when the parameter $a$ is allowed to rise up from zero.
The best fit values for $a$ and $\ell$\/ for the two
samples studied are listed in Table~\ref{tb:summary}.
These best fit values of $\ell$ are consistent with those obtained by
DiCarlo, {\em et al}.~in Ref.~\ref{ref:daved}.

As a test of our assertion that the intrinsic line shape is nominally
a Lorentzian because we have integrated over the $q_\parallel$
direction, we fit these data to the convolution of
the resolution function with a Lorentzian raised
to an arbitrary power~$\eta$.
Similar to a cut off at small $x$,
increasing the value of $\eta$ suppresses the
scattering in the wings.
However, the data is best described by a Lorentzian.
Quantitatively,
for sample~\#1, the best fit value of the exponent was
$\eta \approx 1.09 \pm 0.04$ with
$\chi^2 \approx 13$, roughly twice that obtained using Eq.~(\ref{eq:K_1}),
and excluding a Lorentzian to the $\frac {3}{2}$ line shape.

The principal result of our line shape analysis, the
existence of a second length scale, is not sensitive to
our assumption of a Gaussian resolution function.
This is quite important since
the true resolution function is certainly not expected to be Gaussian.
Both dynamical diffraction\cite{james}
and thermal diffuse scattering\cite{warren}
theory predict that a Bragg peak should have tails which decay only
as $q^{-2}$.
Due to the convolution,
the observed scattering associated with a resolution function with wings
would be enhanced relative to the Gaussian prediction.
But, since the predicted effect is a {\em decrease} in the scattering,
our essential result is insensitive to such tails.

Our line shape analysis of the data demonstrates that
high frequency phase fluctuations are suppressed in NbSe$_{\rm 3}$.
This suppression
is consistent with the predictions of the Ginzburg-Landau field
theory which specifically excludes phase fluctuations
with spatial frequencies greater than $1/\xi$.
The connection between our simple approximate form of $g(x)$ and
the exact formulation can be examined by
equating the Taylor series expansion of
$\sqrt{ a^2 + x^2}/\ell$ with Eq.~(\ref{eq:taylor}),
yielding $a = ( 3 \pi / 4 ) \xi$.
Thus, combining our two data sets, we find that the
amplitude coherence length in the ${\bf a}^*$ direction
is $\xi_{{\bf a}^*} = 34.4 \pm 10.3$\AA.

The observation that high frequency phase fluctuations are suppressed has
implications for our physical picture of the pinning mechanism,
supporting the following
picture which is based on the notion of
a Friedel oscillation\cite{harrison}.
The same divergence of the susceptibility which
drives the Peierls transition
causes the conduction electron density to respond
to a point impurity by creating an oscillating charge
density with wavenumber $2 k_F$.
Of course these oscillations cannot continue to infinite distance in
a real sample.
In general, we expect an exponential decay
of the form $e^{-r/\xi}$ for these oscillations
due to scattering of the conduction electrons by other defects.
The minimum value of the mean free path is,
by definition, the amplitude coherence length $\xi$.
Thus, a pinning site prefers not only a particular value of the phase,
but that the phase be constant over a region
of characteristic size $\xi$.
In contrast to the physical description usually given
for strong pinning
in which the phase gradients are largely confined to a small
region surrounding an individual impurity\cite{tucker},
phase gradients are expelled from the neighborhood of impurity
sites.

Both strong and weak pinning theories are subsumed within
the random field description; however,
the structure of the CDW is determined primarily by the statistics
of the random field, not its intensity.
In our ongoing work studying the kinetics of CDW systems,
the more general random field description is expected to be the most useful.

The authors thank R.E.~Thorne for providing the NbSe$_3$ samples.
This work was supported by Cornell's Materials Science Center (NSF Grant
No.~DMR-88-1858-A02)
and by the NSF (Grant No.~DMR-92-57466).
Additional support was provided by the AT\&T Foundation.
These data were collected on beam line X25 at the
National Synchrotron Light Source,
Brookhaven National Laboratory, which is supported by the
U.S.~Department of Energy, Division of Materials Sciences and Division
of Chemical Sciences (Contract No.~DE-AC02-76CH00016).

\begin{figure}
\caption{
Scan in the
${\bf b}^*$ direction through the [0 $\overline{1+Q_1}$ 0] CDW satellite of
sample~\#1.
Error bars represent counting statistics.
Solid line is the best fit to a Gaussian.
Inset shows ellipse depicting the half-height of the {\em a priori}
resolution function.
The arrows indicate the value predicted for the width.
}
\label{fig:pure_b}
\end{figure}
\begin{figure}
\caption{
Scan in the
${\bf a}^*$ direction through
the [0 $\overline{1+Q_1}$ 0] CDW satellite of
sample~\#1.
The dashed line is the best fit to the convolution of
a Lorentzian with the Gaussian resolution.
The solid line is the best fit to the convolution of
Eq.~(10) with the Gaussian resolution.
The inset shows the central portion of the scan on a linear
scale.
}
\label{fig:pure_a}
\end{figure}
\begin{figure}
\caption{
Scan in the
${\bf a}^*$ direction through
the [0 $\overline{1+Q_1}$ 0] CDW satellite of
sample~\#2.
}
\label{fig:doped_a}
\end{figure}
\begin{table}
\caption{Summary of parameters describing
NbSe$_3$ samples lightly doped with Ta ($r_R \approx 45$).
The errors on the parameters represent 1-$\sigma$ confidence limits of
the nonlinear least squares fit.
}

\begin{tabular}{lr@{$\ \pm \ $}lr@{$\ \pm \ $}l}
\       & \multicolumn{2}{c}{Sample \#1}
& \multicolumn{2}{c}{Sample \#2} \\
\tableline
$a$ & $102$ & $30$ \AA   & $60$ & $38$ \AA \\
$\ell_{{\bf a}^*}$ & $3\:275$ & $175$ \AA      & $3\:050$ & $400$ \AA \\
W$_{\rm R}$ (HWHM)& \multicolumn{2}{c}{$3.3 \times 10^{-4}$~\AA$^{-1}$} &
\multicolumn{2}{c}{$1.32 \times 10^{-3}$~\AA$^{-1}$} \\
I$_{\rm BG}$& \multicolumn{2}{c}{0.61 counts/sec} &
\multicolumn{2}{c}{0.24 counts/sec} \\
\end{tabular}

\label{tb:summary}
\end{table}


\begin{thebibliography}{99}
\bibitem{comment} References \ref{ref:abernathy} and \ref{ref:mochrie}
stressed that it was important to keep $g (0) = 0$; however, the present
work clearly demonstrates that it is the functional form,
not the absolute value, which must be maintained.
An incorrect absolute value can simply rescaled away
by a constant multiplicative factor.
An incorrect functional form always produces an incorrect line shape.
\end{thebibliography}
\end{document}